\documentclass{aa}

\usepackage{graphicx}
\usepackage{epsfig}
\usepackage{natbib}
\usepackage[varg]{txfonts}
\usepackage{hyperref}

\newcommand{\COBOLD}{{\tt CO$^{\tt 5}$BOLD}}

\newcommand{\LHD}{{\tt LHD}}
\newcommand{\LINFOR}{{\tt Linfor3D}}
\newcommand{\ATLAS}{{\tt ATLAS9}}
\newcommand{\Synthe}{{\tt SYNTHE}}

\newcommand{\MARCS}{{\tt MARCS}}
\newcommand{\teff}{\ensuremath{T_{\mathrm{eff}}}}
\newcommand{\logg}{\ensuremath{\mathrm{log}\,g}}
\newcommand{\MoH}{\ensuremath{\left[\mathrm{M}/\mathrm{H}\right]}}
\newcommand{\FeH}{\ensuremath{\left[\mathrm{Fe}/\mathrm{H}\right]}}

\newcommand{\OFe}{\ensuremath{\left[\mathrm{O}/\mathrm{Fe}\right]}}
\newcommand{\mlp}{\ensuremath{\alpha_{\mathrm{MLT}}}}
\newcommand{\xtmean}[1]{\ensuremath{\left\langle #1\right\rangle}}

\begin{document}

\title{Three-dimensional hydrodynamical \COBOLD\ model atmospheres of red giant stars
}
\subtitle{IV. Oxygen diagnostics in extremely metal-poor red giants with infrared OH lines\thanks{Based on observations obtained at the European Southern Observatory (ESO) Very Large Telescope (VLT) at Paranal Observatory, Chile (observing programme 089.D-0079).}
}

\author{
       V. Dobrovolskas
       \inst{1}
       \and
       A. Ku\v{c}inskas
       \inst{1,2}
       \and
       P.~Bonifacio
       \inst{3}
       \and
       E.~Caffau
       \inst{3,4}
       \and
       H.-G.~Ludwig
       \inst{4,3}
       \and
       M.~Steffen
       \inst{5,3}
       \and
       M.~Spite
       \inst{3}
       }

 \institute
       {Institute of Theoretical Physics and Astronomy, Vilnius University, Go\v{s}tauto 12, Vilnius, LT-01108, Lithuania\\
       \email{vidas.dobrovolskas@tfai.vu.lt}
       \and
       Vilnius University Astronomical Observatory, M. K. \v{C}iurlionio 29, Vilnius, LT-03100, Lithuania
       \and
       GEPI, Observatoire de Paris, CNRS, Universit\'{e} Paris Diderot, Place Jules Janssen, 92190 Meudon, France
       \and
       Landessternwarte -- Zentrum f\"ur Astronomie der Universit\"at Heidelberg, K\"{o}nigstuhl 12, D-69117 Heidelberg, Germany
       \and
       Leibniz-Institut f\"ur Astrophysik Potsdam, An der Sternwarte 16, D-14482 Potsdam, Germany\\
       }

\date{Received ; accepted }

\abstract
{Although oxygen is an important tracer of Galactic chemical evolution, measurements of its abundance in the atmospheres of the oldest Galactic stars are still scarce and rather imprecise. This is mainly because only a few spectral lines are available for the abundance diagnostics. At the lowest end of the metallicity scale, oxygen can only be measured in giant stars and in most of cases such measurements rely on a single forbidden \ensuremath{\left[\ion{O}{i}\right]} 630~nm line that is very weak and frequently blended with telluric lines. Although molecular OH lines located in the ultraviolet and infrared could also be used for the diagnostics, oxygen abundances obtained from the OH lines and the \ensuremath{\left[\ion{O}{i}\right]} 630~nm line are usually discrepant to a level of $\sim0.3-0.4$~dex.}
{We study the influence of convection on the formation of the infrared (IR) OH lines and the forbidden \ensuremath{\left[\ion{O}{i}\right]} 630~nm line in the atmospheres of extremely metal-poor (EMP) red giant stars. Our ultimate goal is to clarify whether a realistic treatment of convection with state-of-the-art 3D hydrodynamical model atmospheres may help to bring the oxygen abundances obtained using the two indicators into closer agreement.
}
{We used high-resolution ($R=50\,000$) and high signal-to-noise ratio ($S/N\approx200-600$) spectra of four EMP red giant stars obtained with the VLT~CRIRES spectrograph. For each EMP star, 4--14 IR OH vibrational-rotational lines located in the spectral range of $1514-1548$ and $1595-1632$~nm were used to determine oxygen abundances by employing standard 1D local thermodynamic equilibrium (LTE) abundance analysis methodology. We then corrected the 1D~LTE abundances obtained from each individual OH line for the 3D hydrodynamical effects, which was done by applying 3D--1D~LTE abundance corrections that were determined using 3D hydrodynamical \COBOLD\ and 1D hydrostatic \LHD\ model atmospheres.}
{We find that the influence of convection on the formation of \ensuremath{\left[\ion{O}{i}\right]} 630~nm line in the atmospheres of EMP giants studied here is minor, which leads to very small 3D--1D abundance corrections ($\Delta_{\rm 3D-1D}\le -0.01$~dex). On the contrary, IR OH lines are strongly affected by convection and thus the abundance corrections for these lines are significant, $\Delta_{\rm 3D-1D}\approx-0.2 \dots -0.3$~dex. These abundance corrections do indeed bring the 1D~LTE oxygen abundances of EMP red giants obtained using IR OH lines into better agreement with those determined from the \ensuremath{\left[\ion{O}{i}\right]} 630~nm line. Since in the EMP red giants IR OH lines are typically at least a factor of two stronger than the \ensuremath{\left[\ion{O}{i}\right]} line, OH lines may be useful indicators of oxygen abundances in the EMP stars, provided that the analysis is based on 3D hydrodynamical model atmospheres.
}
{}

\authorrunning{Dobrovolskas et al.}
\titlerunning{Oxygen diagnostics with infrared OH lines}

\maketitle

\section{Introduction}

Oxygen is the most abundant element in the Universe, after hydrogen and helium, and is the main product of the Type~II SNe
\citep[see e.g.][and references therein]{Arnett}. Massive stars have lifetimes of the order of 1$0^6$ to $10^7$ years and end their lives as
supernovae, releasing  oxygen  into the interstellar medium \citep[see e.g.][and references therein]{Matteucci}. As a first approximation it may be expected that the oxygen abundance in extremely metal-poor (EMP) stars correlates with their age, the most oxygen-poor stars being the oldest. On closer consideration it becomes clear that the oxygen abundance at any given time also depends on the star formation rate. In a complex system, like our Galaxy, that may host populations of different origins (perhaps some accreted by dwarf satellite galaxies) the evolution of the O/Fe ratio with time depends on the details of the chemical evolution of the different populations. The abundance ratio of oxygen to other elements, and in particular to iron, places strong constraints on  proposed scenarios of Galactic chemical evolution. The precise knowledge of oxygen abundances in the oldest Galactic populations is important since it allows us to discriminate among different models \citep[see ][chapter 5.3, for a discussion]{Matteucci}.

In spite of these virtues, the oxygen abundance is actually very difficult to measure in the metal-poor stellar atmospheres: the only indicators available in the EMP stars are (i) the forbidden \ensuremath{\left[\ion{O}{i}\right]} line at 630 nm for giants, (ii) the permitted $\ion{O}{i}$ IR triplet for dwarfs ($777.2 - 777.5$~nm), and (iii) OH lines in the UV and IR, for both giants and dwarfs. Unfortunately, these different indicators often provide discrepant oxygen abundances, which may be due to non-local thermodynamic equilibrium (NLTE) effects (especially for the IR triplet), 3D hydrodynamical effects (especially for the OH lines but also for the triplet), incorrectly determined atmospheric parameters (e.g. \ensuremath{\left[\ion{O}{i}\right]} is very sensitive to the surface gravity; see e.g. \citealt{CAB01,ANA06}), and so on. Consequently, reliable oxygen abundances of the EMP stars are scarce: the number of objects in which they have been measured so far is small while the measurements themselves are frequently not sufficiently reliable. This has led different groups to claim the existence of different trends of \OFe\ with decreasing metallicity \citep[see e.g.][]{BKD99,IRG01,ACN04}, which makes discriminating between the various proposed scenarios of Galactic oxygen evolution very difficult. Clearly, such a situation is unsatisfactory and calls for the re-assessment of oxygen abundances in EMP stars.

It should be noted that oxygen abundances in the EMP giants are nearly always obtained from the measurements of a single forbidden \ensuremath{\left[\ion{O}{i}\right]} 630~nm line. One reason why OH lines have rarely been used for the abundance determinations so far \citep{MBS01,MB02,BMS03} is that the oxygen abundances obtained from these lines are significantly different from those determined using the \ensuremath{\left[\ion{O}{i}\right]} line \citep[e.g.][]{BMS03}. Part of the problem may be related to the \ensuremath{\left[\ion{O}{i}\right]} line measurements, because this line is very weak and often blended with telluric lines. On the other hand, molecular lines may be prone to 3D hydrodynamical effects, such as horizontal temperature fluctuations, which may lead to substantial differences in the oxygen abundances derived with the 3D hydrodynamical and classical 1D hydrostatic models \citep[see e.g.][]{CAT07,DKS13}. It may thus be reasonable to anticipate that oxygen abundances obtained from OH lines and those determined using the \ensuremath{\left[\ion{O}{i}\right]} line could be brought closer if these effects are properly taken into account in the model atmospheres used in the abundance determination. If this were proved to be true, OH lines could be used as independent valuable indicator of the oxygen abundance, with the advantage that (i) OH lines are numerous in the infrared and thus one would not need to rely on a single line for measuring the oxygen abundance, as in the case with the forbidden \ensuremath{\left[\ion{O}{i}\right]} 630~nm line; and (ii) infrared (IR) OH lines are at least a factor of 2 stronger than the \ensuremath{\left[\ion{O}{i}\right]} 630~nm line is, which makes them easier to measure and may lead to a better accuracy in the determined oxygen abundances.

In this pilot study we therefore investigate whether such considerations are indeed valid in the case of EMP red giant stars. For this purpose, we used high-resolution and high signal-to-noise spectra of four EMP giants obtained with the CRIRES spectrograph at the VLT~UT1 telescope to determine 1D~LTE oxygen abundances using IR~OH lines. Then, 3D hydrodynamical \COBOLD\ \citep{FSL12} and 1D hydrostatic \LHD\ \citep{CL07} model atmospheres were used in order to assess the role and influence of convection on the formation of IR~OH lines and the forbidden \ensuremath{\left[\ion{O}{i}\right]} line, as well as to determine the 3D-corrected abundances of oxygen from IR~OH lines. Finally, we discuss the possible role of various other effects that may influence the formation of IR~OH lines and that may need to be taken into account in the abundance determinations using IR~OH lines.

\section{Methodology\label{sect:methodol}}

\subsection{Observations and data reduction\label{sect:obs-data}}

Since IR~OH lines are very weak in EMP giants, one needs to have access to spectra with very high $S/N$ ratios in order to obtain reliable oxygen abundances. For this reason, we limited our target list to objects with $V < 12.0$; these are the brightest EMP giants in the list of \citet[][]{CDS04}. This has allowed us to keep the observing time request within reasonable limits even in the case of the faintest star, CS~31082-001; we note that except for HD~122563, no EMP giants have been observed with CRIRES so far. The list of our EMP targets is provided in Table~\ref{table:obs-star}.

High-resolution $H$-band near-infrared spectra of the four EMP giants were obtained in  service mode with the CRIRES spectrograph \citep{KBB04} located at the VLT UT1 telescope (programme ID 089.D-0079(A), PI: A.~Ku\v{c}inskas). We used the entrance slit width of 0.4~arcsec for our observations and did not use the adaptive optics system. The spectra were acquired in ABBA nodding pattern for dark current and sky emission removal, using two CRIRES settings: order 37 with the central wavelength of 1536.4~nm and a spectral range of $1513.4-1548.6$~nm; and order 35 with the central wavelength of 1619.4~nm and a spectral range of $1594.9-1632.5$~nm. The targets were observed with a spectral resolution of $R = 50\,000$ and signal-to-noise ratio per pixel $S/N\approx200-600$. Standard steps involved in the reduction of the raw spectra (bad pixel correction, detector non-linearity correction, flat-fielding, and wavelength calibration) were performed using the CRIRES pipeline (v.2.3.1) in the command-line (EsoRex) mode.

Unfortunately, the wavelength calibration solution provided by the CRIRES pipeline had systematic shifts which made spectral line identification difficult. Therefore, we performed a custom wavelength calibration using telluric lines present in the observed spectral range. For this purpose, we computed the synthetic telluric spectrum using the TAPAS online service \footnote{\url{http://ether.ipsl.jussieu.fr/tapas/}} \citep{BLF14} for the conditions matching those of our observations (location, date, time, and air-mass). We did not attempt to clean stellar spectra from the telluric lines, thus for further spectroscopic analysis we used only those OH lines which were free from telluric contamination. In fact, we observed a telluric star HD~24388, but because of a significantly lower S/N ratio this spectrum was not used to correct the target star spectra for the telluric features. Instead, we chose the synthetic telluric spectrum computed using TAPAS to identify telluric lines in our CRIRES spectra and to avoid them in further abundance analysis. Continuum normalization of the observed CRIRES spectra was made using the \texttt{DECH20T} package \citep{galazutdinov}.

\begin{table*}[tb]
\caption{List of EMP giants observed with CRIRES.}
\label{table:obs-star}
\centering
\begin{tabular}{lccrrrccccc}
\hline\hline
Star          & $\alpha(\rm J2000)$  & $\delta(\rm J2000)$ & $V$, & $H$, & $t_{\rm exp}$,   &  $S/N$  &  \teff,  &  \logg  & $\xi$       & \FeH     \\
              &                      &                     &(mag) &(mag) &(s)               &         & (K)      & [cgs]   &(km\,s$^{-1}$)&         \\
\hline
 HD~122563    & 14 02 31.85          & $+09\,41\,09.9$     & 6.2  & 3.8  & 35   &  560    &  4598    &  1.60   &  2.0  &  $-2.82$  \\
 HD~186478    & 19 45 14.14          & $-17\,29\,27.1$     & 9.2  & 6.6  & 270  &  440    &  4700    &  1.30   &  2.0  &  $-2.59$  \\
 BD~--18:5550 & 19 58 49.74          & $-18\,12\,11.1$     & 9.4  & 6.7  & 270  &  600    &  4750    &  1.40   &  1.8  &  $-3.06$  \\
 CS~31082-001 & 01 29 31.13          & $-16\,00\,45.4$     & 11.3 & 9.6  & 1500 &  190    &  4825    &  1.50   &  1.8  &  $-2.91$  \\
\hline
\end{tabular}
\begin{list}{}{}
\item
Note: atmospheric parameters and iron abundances of the target stars are from \citet{SCP05}, except for HD~122563 which are from \citet{CTB12}. Microturbulence velocities, $\xi$, are from \citet{SCP05}, $V$ and $H$ magnitudes are from SIMBAD.
\end{list}
\end{table*}

\begin{table*}[tb]
\caption{Oxygen abundances in EMP giants.}
\label{table:o-abnd-tab}
\centering
\begin{tabular}{lcccccc}
\hline\hline
Star         & $A({\rm O})^{\rm 1D\,LTE, a}$  & $\sigma A({\rm O})^{\rm 1D\,LTE, a}$ &  $ A({\rm O})^{\rm 3D\,LTE, a} $ & $\sigma A({\rm O})^{\rm 3D\,LTE, a}$ & $A({\rm O})^{\rm 1D\,LTE, b}$ & $\sigma A({\rm O})^{\rm 1D\,LTE, b} $\\
             & OH                  &  OH    &  OH                      &  OH  &  \ensuremath{\left[\ion{O}{i}\right]}                                & \ensuremath{\left[\ion{O}{i}\right]} \\
\hline
HD~122563    & 6.63  &  0.10  &   6.39 &  0.11  &  6.57  &  0.15  \\
HD~186478    & 7.11  &  0.09  &   6.85 &  0.08  &  6.93  &  0.14  \\
BD~--18:5550 & 6.73  &  0.13  &   6.48 &  0.15  &  6.13  &  0.26  \\
CS~31082-001 & 6.98  &  0.13  &   6.73 &  0.10  &  6.46  &  0.20  \\
\hline
\end{tabular}
\begin{list}{}{}
\item[$^{\mathrm{a}}$] determined in this work; $^{\mathrm{b}}$ from \citet{CDS04}
\end{list}
\end{table*}

\subsection{Model atmospheres \label{sect:models}}

Three types of model atmospheres were employed in this study. First, 1D~LTE oxygen abundances were determined by using model atmospheres computed with the \ATLAS\ code \citep[][]{kurucz93}. We used the \ATLAS\ version that was ported to Linux by \citet[][]{sbordone04} and \citet[][]{sbordone05}. The models were calculated using opacity distribution functions (ODFNEW) from \citet[][]{castelli03}; the latter were computed with the microturbulence velocity of $\xi=2\,{\rm km\,s^{-1}}$. All \ATLAS\ model atmospheres were calculated using the mixing length parameter $\alpha_{\rm MLT}=1.25$, with overshooting switched off.

In addition, 3D hydrodynamical and 1D hydrostatic model atmospheres were used for computing 3D--1D abundance corrections (see Sect.~\ref{sect:3d-abnd-corr}). For this purpose we employed a single 3D hydrodynamical model atmosphere computed using the \COBOLD\ model atmosphere package \citep{FSL12} with the following atmospheric parameters: $\teff=4595 {~\rm K}, \logg=1.6, \MoH=-2.5$ (Klevas et al., in prep.). Our choice was determined by the fact that 3D hydrodynamical models are rather expensive to compute and thus they are not yet available for all combinations of atmospheric parameters. Nevertheless, atmospheric parameters of this particular model atmosphere are sufficiently similar to those of EMP giants studied here: even in the most extreme cases the differences do not exceed $\Delta \teff=230$~K, $\Delta \log g = 0.3$, and $\Delta \MoH = 0.5$, while generally they are even smaller (see Table~\ref{table:obs-star}).

The 3D hydrodynamical model atmosphere was computed in Cartesian geometry with a spatial resolution of $220\times220\times280$ cells in the $x,y,z$ directions, respectively, which corresponds to spatial dimensions of $3.85\times3.85\times2.21$ Gm. The horizontal size of the model was large enough to accommodate $\approx 10$ granules which allowed convective dynamics to be reliably captured in the stellar atmosphere \citep[see][]{CAT07,LK12,KSL13}. We used opacities from the \MARCS\ model atmosphere package, which -- in order to make the model computations faster -- were grouped into six opacity bins using opacity binning \citep{nordlund82,ludwig92,LJS94,VBS04}. Solar chemical composition from \citet{GS98} with $\alpha$-element enhancement of +0.4~dex was used in the model calculations. For CNO elements, we used solar values of $A({\rm C}) = 8.41$, $A({\rm N}) = 7.8$, and $A({\rm O}) = 8.66$ taken from \citet{AGS05}, which were then scaled down to the metallicity of our 3D hydrodynamical model atmosphere with $\alpha$-element enhancement of +0.4~dex.

In order to speed up the 3D spectral line synthesis calculations, we selected 20 3D model structures computed at different instants in time (snapshots). When making the snapshot selection we tried to ensure that the average properties of the selected snapshot subsample would remain similar to the average properties of the full 3D model simulation run \citep[see e.g.][for the details on the snapshot selection procedure]{KSL13}. This snapshot subsample was used in subsequent 3D spectral synthesis computations.

The 20-snapshot ensemble was also used to compute average $\xtmean{\mbox{3D}}$ model, which was obtained by averaging the 3D structures in the 20-snapshot subsample on surfaces of equal optical depth. As has been already discussed in our previous studies, such average $\xtmean{\mbox{3D}}$ models are useful in evaluating the role of horizontal temperature fluctuations in the spectral line formation. Since the average $\xtmean{\mbox{3D}}$ models do not contain information about the horizontal fluctuations of thermodynamic properties, the 3D--$\xtmean{\mbox{3D}}$ differences provide important information about the role of these fluctuations in the spectral line formation \citep[see e.g.][]{DKS13,KSL13}.

The 1D hydrostatic model atmospheres that we used in the evaluation of 3D--1D abundance corrections were computed using \LHD\ model atmosphere package \citep[][]{CLS08}. It is important to note that 3D hydrodynamical and 1D hydrostatic \LHD\ model atmospheres employed in the computation of abundance corrections shared identical atmospheric parameters, chemical composition, opacities, and equation of state. This allowed us to make a strictly differential comparison and to estimate the impact of convection on the formation of spectral lines. Therefore, we used the (differential) 3D--1D abundance corrections obtained in this way to correct the abundances determined using the \ATLAS\ hydrostatic model atmosphere computed with different opacities. The 1D \LHD\ models were calculated with the mixing-length parameter $\mlp=1.0$.

We note that effects of sphericity may in fact affect the structure of the model atmosphere, both in 1D \citep[see e.g.][]{HE06} and,  possibly, in 3D as well. To assess the size of these effects in 1D, we performed a comparison of a spherically symmetric and a plane-parallel model atmospheres with $\teff=4500 {~\rm K}, \logg=1.5, \MoH=-2.5$. The choice was dictated by the available models, and in fact the spherical model was a \MARCS\ model, the plane-parallel an \ATLAS\ model. The spherical model had a radius of 29\,R$_\odot$ and an atmospheric extension of 0.5\,R$_\odot$. We synthesized a few typical OH lines and the \ensuremath{\left[\ion{O}{i}\right]} line with \LINFOR, under the assumption of Cartesian geometry. Despite this approximation, according to \citet{HE06}, the main effect should be captured by this procedure. We found a slight strengthening of the \ensuremath{\left[\ion{O}{i}\right]} line ($\approx 0.01$\,dex), and a modest strengthening of the OH lines ($\approx 0.06$\,dex) in the spherical versus the plane-parallel model. From this we conclude that sphericity effects are indeed present but on a level significantly smaller than the 3D effects.

Unfortunately, we are currently not able to assess the size of sphericity effects on the 3D hydrodynamical model structures because the computation of spherical 3D hydrodynamical \COBOLD\ models is not yet feasible. Nevertheless, we do not expect that sphericity effects play a major role for the 3D--1D abundance corrections discussed here since the modest extension of the atmosphere is unlikely to change the granular dynamics noticeably. The overall effect of the geometry should largely cancel out because of the differential nature of the 3D--1D corrections.

\section{Results \label{sect:results}}

\subsection{1D~LTE oxygen abundances\label{sect:1d-abnd}}

To determine oxygen abundances in the target EMP giants, we used infrared OH vibrational-rotational lines ($X^{2}\Pi$) from the first-overtone sequence ($\Delta v = 2$) located between $1514-1548$ and $1595-1632$~nm (see Table~\ref{table:line-list} and Figs.~\ref{fig:hd122563-spect}-\ref{fig:cs31082-spect} in Online Material). We note that the spin-orbit interaction gives rise to  two spin-split substates: $^2\Pi_{1/2}$ and $^2\Pi_{3/2}$. All the lines that we consider are from the P-branch ($\Delta J = -1$), the rotational levels are subject to $\Lambda$-doubling, i.e. splitting that is due to  perturbation by the rotational levels of other electronic states \citep[see][]{coxonfoster}. The equivalent widths of the observed lines were measured using the \texttt{splot} task in \texttt{IRAF}\footnote{IRAF is distributed by the National Optical Astronomy Observatories, which are operated by the Association of Universities for Research in Astronomy, Inc., under cooperative agreement with the National Science Foundation.} and are provided in Table~\ref{table:line-list}. Although some of the OH lines used in the abundance determination are contaminated by telluric lines, these telluric lines are in fact very weak. Therefore, we did not correct for their influence when measuring the equivalent widths of OH lines. We note that all observed targets have known \ensuremath{\left[\ion{O}{i}\right]} line strengths, as well as oxygen abundances determined using this line (provided in Table~\ref{table:o-abnd-tab}). As expected, in most cases the IR~OH lines are significantly stronger than the forbidden \ensuremath{\left[\ion{O}{i}\right]} line at 630~nm.

We also note that although there are CN lines in the spectral range investigated in this study, they are very weak ($\le 0.05$\% depth from continuum) and their influence on the oxygen abundance determination is negligible, even in cases where they blend with OH lines. All the stars investigated in this study have a normal or low carbon abundance of [C/Fe] < 0.2 \citep{takeda13}, which leads to very weak CN lines in their IR spectra.

Oxygen 1D LTE abundances were determined using the curve-of-growth method, employing \ATLAS\ model atmospheres computed as described in Sect.~\ref{sect:models} above. The \ATLAS\ models were calculated using the atmospheric parameters of the individual EMP giants that were taken from the literature and are provided in Table~\ref{table:obs-star}. Synthetic spectra used in the abundance determination were computed using the Linux version of the \Synthe\ spectral synthesis package \citep{sbordone04,sbordone05}. We then measured the equivalent widths of the OH lines in the synthetic spectra using the \texttt{splot} task in \texttt{IRAF}, i.e. using the same methodology as for the observed lines. Finally, 1D~LTE oxygen abundances were determined by interpolating between the curves of growth at the equivalent widths of IR~OH lines that we measured in each EMP giant (Table~\ref{table:o-abnd-tab}).

We estimate that the uncertainty in the determined oxygen abundance arising due to the uncertainty in the continuum level determination (which we estimate to be $\pm0.2$~\%) is $\pm0.06$~dex. The uncertainty in the measured oxygen abundance provided in Table~\ref{table:o-abnd-tab} is a square root of the sum of (i) the dispersion due to line-to-line variations of the determined oxygen abundance for each EMP star and (ii) the squared uncertainty due to continuum placement.

Many OH lines studied here form from transitions that involve rotational levels which, because of perturbations by other electronic states, are split into two sub-levels ($\Lambda$-doubling). Therefore, the two lines arising because of such transitions should give identical oxygen abundances. Since in reality this, apparently, is not the case, the difference between abundances obtained from the two individual components may provide additional constraints on the size of the uncertainties in the determined oxygen abundances. In our case, we find that the largest difference in the oxygen abundance obtained from two lambda-doubling components is $\sim0.12$~dex, which is comparable with the uncertainties provided in Table~\ref{table:o-abnd-tab}.

\begin{figure}[tb]
\centering
\includegraphics[width=\columnwidth]{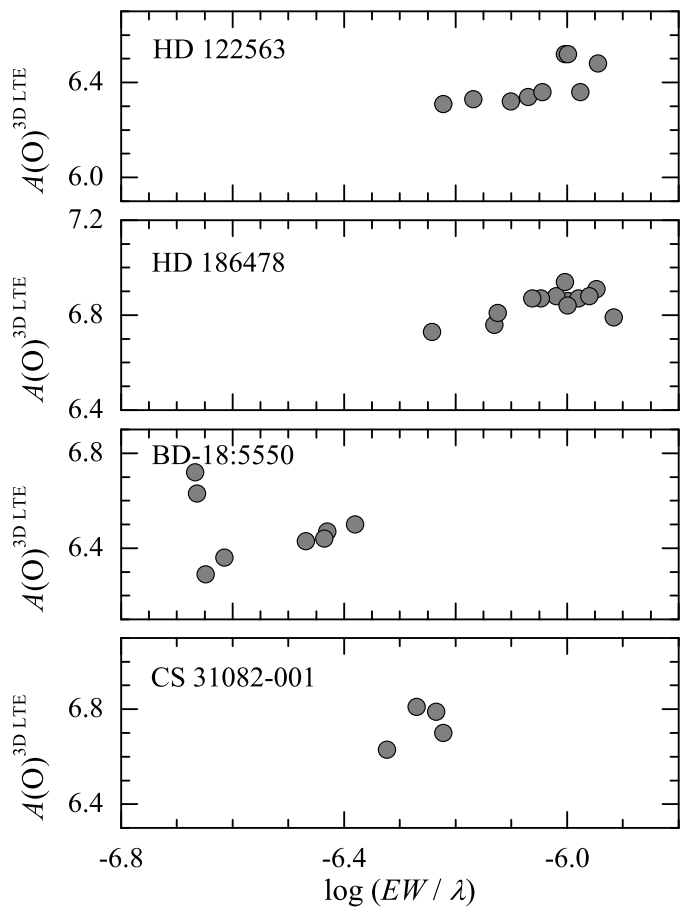}
\caption{3D~LTE oxygen abundance in HD 122563, HD 186478, BD --18:5550, and CS 31082-001 (top to bottom), as determined from IR~OH lines, shown as the function of reduced equivalent width.}
\label{fig:o-abnd-1d}
\end{figure}

\begin{figure}[tb]
\centering
\includegraphics[width=\columnwidth]{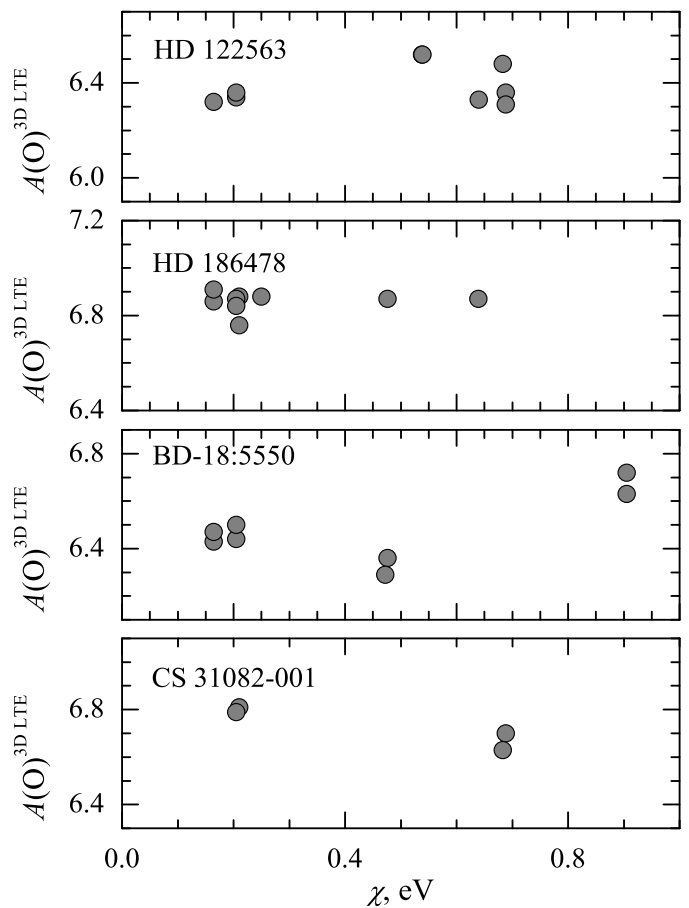}
\caption{3D~LTE oxygen abundance in HD 122563, HD 186478, BD --18:5550, and CS 31082-001 (top to bottom), as determined from IR~OH lines, shown as the function of excitation potential.}
\label{fig:o-abnd-1d_exct}
\end{figure}

\subsection{3D--1D abundance corrections and 3D~LTE oxygen abundances \label{sect:3d-abnd-corr}}

\begin{figure}[tb]
\centering
\includegraphics[width=\columnwidth]{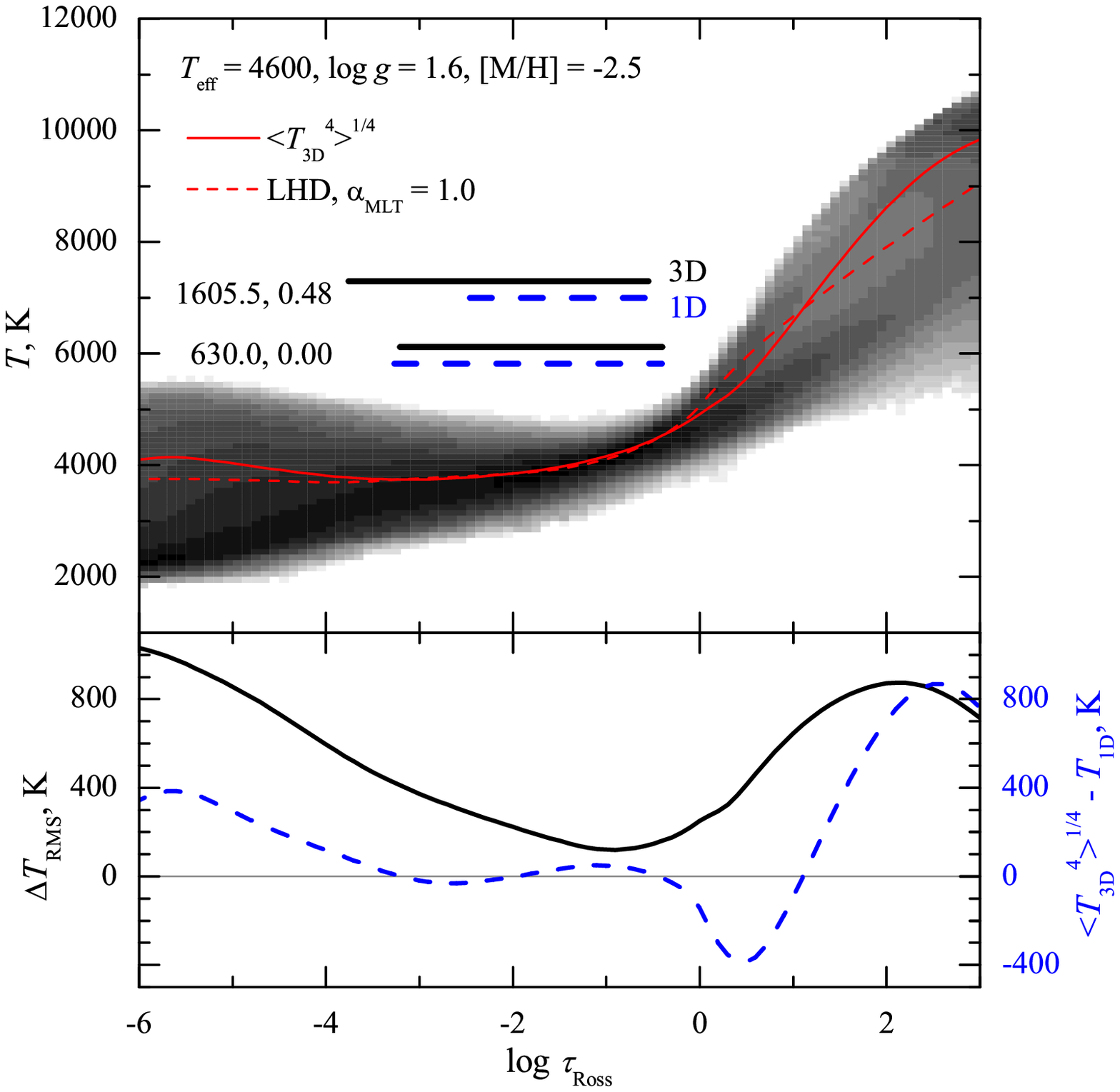}
 \caption{\textbf{Top:} Temperature stratifications in the 3D hydrodynamical (grey scales indicating the temperature probability density), average $\xtmean{\mbox{3D}}$ (solid red line), and 1D \LHD\ (dashed red line) model atmospheres computed with $\teff=4600\, {\rm K}, \logg=1.6, \MoH=-2.5$. Horizontal bars show the optical depth intervals where 90\% of the OH line ($\lambda=1605.5\, {\rm nm}, \chi=0.48\, {\rm eV}$) and \ensuremath{\left[\ion{O}{i}\right]} line ($\lambda=630.0\, {\rm nm}, \chi=0.00\, {\rm eV}$) equivalent width is formed (i.e. 5\% to 95\%): black and dashed blue bars correspond to the line forming regions in the full 3D and 1D model atmospheres (the equivalent widths of OH and \ensuremath{\left[\ion{O}{i}\right]} lines are 1.6~pm and 0.4~pm). \textbf{Bottom:} RMS value of horizontal temperature fluctuations, $\Delta T_{\rm RMS}$, in the 3D model (black line) and temperature difference between the average $\xtmean{\mbox{3D}}$ and 1D models (dashed blue line).}
 \label{fig:temp-strat}
\end{figure}

\begin{figure}[tb]
\centering
\includegraphics[width=\columnwidth]{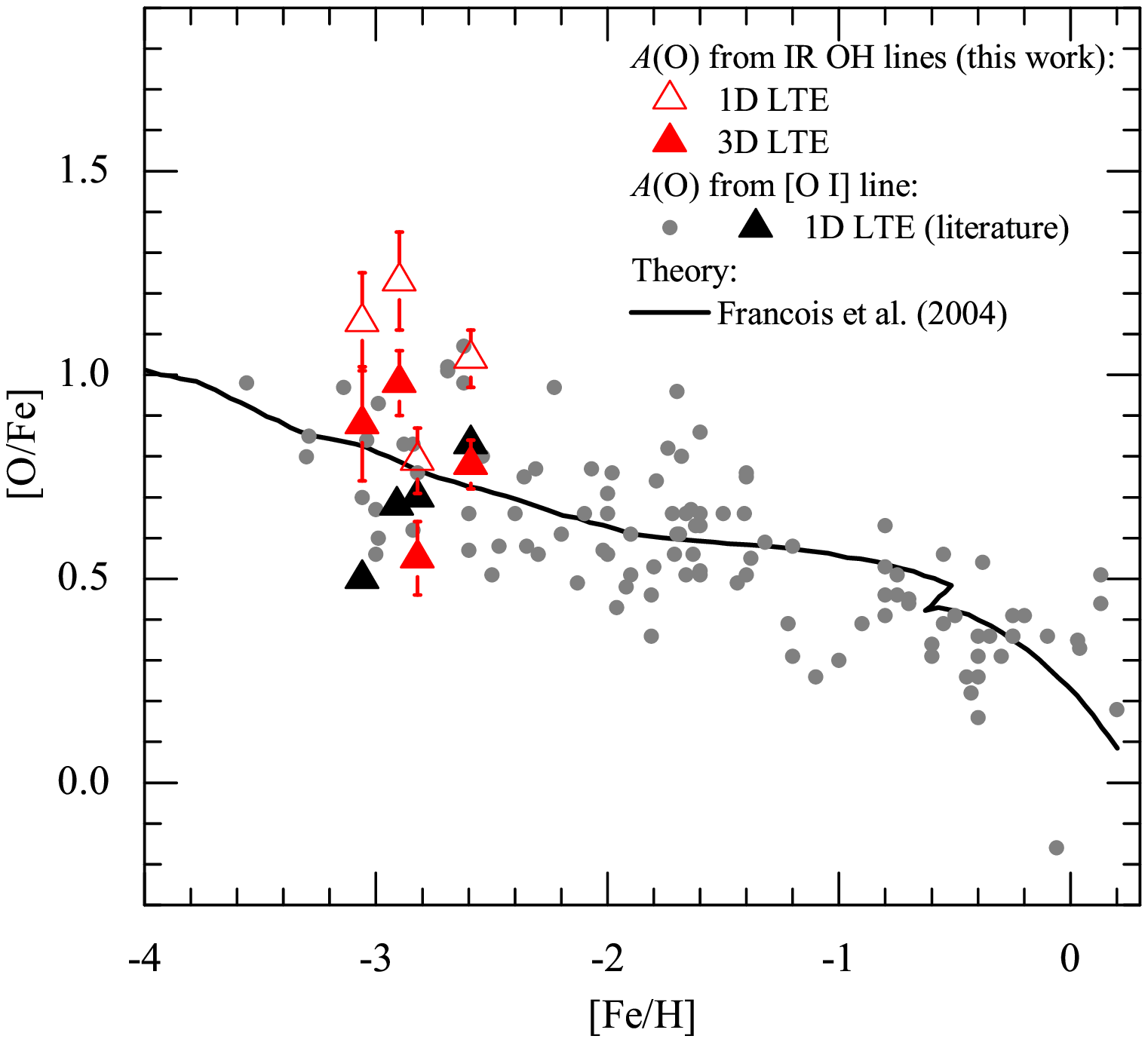}
\caption{Oxygen-to-iron ratios in Galactic EMP stars plotted versus \FeH\ ratio. Abundances of EMP giants studied here are marked by triangles: open red triangles -- oxygen abundances in the programme stars determined determined in this work with classical 1D model atmospheres; filled red triangles -- oxygen abundances corrected for 3D hydrodynamical effects. Solid grey circles show oxygen abundances in red giants, subgiants, and main sequence stars determined using the forbidden \ensuremath{\left[\ion{O}{i}\right]} 630~nm line \citep{GO86,BE89,KSL92,barbuy88,SKP91,FK99,CAB01,CSB02,CDS04}. Black triangles highlight oxygen abundances determined from \ensuremath{\left[\ion{O}{i}\right]} 630~nm line in \citet{CDS04} for the same stars as studied here. The solid line is the evolutionary model of the oxygen abundance from \citet{FMC04}. We note that \ensuremath{\left[\ion{O}{i}\right]} abundances from \citet[][]{CDS04} and the theoretical model from \citet{FMC04} were shifted to the reference scale used in our study where solar oxygen abundance is $A({\rm O}) = 8.66$.}
\label{fig:ofe-evol}
\end{figure}

The 3D~LTE abundances of oxygen were obtained by using 3D--1D abundance corrections. The 3D--1D abundance correction, $\Delta_{\rm3D-1D}=A({\rm X_{i}})_{\rm 3D} - A({\rm X_{i}})_{\rm 1D}$, is the difference in the abundance of element ${\rm X_{i}}$ that would be obtained at a fixed equivalent width of a given spectral line with 3D hydrodynamical and 1D hydrostatic model atmospheres, respectively. In this work, abundance corrections were obtained using the curves of growth computed with 3D hydrodynamical \COBOLD\ and 1D hydrostatic \LHD\ model atmospheres. Synthetic spectral line profiles were calculated under the assumption of LTE in 3D and 1D, in both cases using the \LINFOR\footnote{http://www.aip.de/$\sim$mst/Linfor3D/linfor\_3D\_manual.pdf} spectral synthesis package. Abundance corrections were then computed by measuring the difference between the curves of growth obtained with the 3D hydrodynamical and 1D hydrostatic model atmospheres, respectively.

The 3D--1D abundance corrections were determined for all IR~OH lines used in this work, as well as for the forbidden \ensuremath{\left[\ion{O}{i}\right]} 630~nm line, using the line parameters provided in Table~\ref{table:line-list} and Tables~\ref{table:dabu-hd122563}-\ref{table:dabu-cs31082}. Apart from $\Delta_{\rm3D-1D}$ correction, we also computed two additional corrections, $\Delta_{\rm 3D-\langle3D\rangle}=A({\rm X_{i}})_{\rm 3D}-A({\rm X_{i}})_{\rm \langle3D\rangle}$, and $\Delta_{\rm \langle3D\rangle-1D}=A({\rm X_{i}})_{\rm \langle3D\rangle}-A({\rm X_{i}})_{\rm 1D}$. The former is useful in assessing the importance of horizontal fluctuations of thermodynamic and hydrodynamical quantities in 3D hydrodynamical model atmospheres, while the latter allows the role of differences to be estimated between the temperature profiles in the average $\xtmean{\mbox{3D}}$ and 1D model atmospheres (see Sect.~\ref{sect:discussion}). Abundance corrections for IR~OH lines used in this work with all EMP giants are provided in Tables~\ref{table:dabu-hd122563}--\ref{table:dabu-cs31082}. In these tables we also provide abundance corrections for the forbidden \ensuremath{\left[\ion{O}{i}\right]} 630~nm line.

The 3D~LTE oxygen abundances were obtained by adding $\Delta_{\rm3D-1D}$ correction to 1D~LTE oxygen abundances obtained in Sect.~\ref{sect:1d-abnd} above. Although we are clearly aware of the many limitations of such approach (see Sect.~\ref{sect:discussion} for discussion), we nevertheless think that the 3D-corrected oxygen abundances, when compared to those derived in 1D~LTE, may provide important qualitative information for the assessment of the role of convection in the spectral line formation taking place in the atmospheres of EMP giants. The obtained 3D~LTE abundances are listed in Table~\ref{table:o-abnd-tab} for all EMP giants studied here. The obtained abundances are plotted in Fig.~\ref{fig:o-abnd-1d} versus the line reduced equivalent width, $\log(EW/\lambda)$, and in Fig.~\ref{fig:o-abnd-1d_exct} versus the line excitation potential, $\chi$, for all four EMP stars studied here.

We point out that our 3D--1D oxygen abundance corrections for different programme stars were computed using a single 3D hydrodynamical model atmosphere, whereas the atmospheric parameters of the programme stars differ by up to $+230$~K and $-0.3$~dex in \teff\ and $\log g$ with respect to those of the 3D model atmosphere. With the 1D \ATLAS\ models, this would amount to differences in the determined oxygen abundance of up to $+0.65$~dex and $+0.12$~dex, respectively. These numbers are very similar to those obtained by \citet[][]{BMS03}. Unfortunately, our 3D hydrodynamical model grid is too sparse to estimate the size of temperature and gravity sensitivity of the 3D--1D abundance corrections. Nevertheless, our crude estimate is that an increase in the effective temperature by $+230$~K (from $\teff=4600$~K) would change the full $\Delta_{\rm3D-1D}$ abundance correction by $-0.06 \dots -0.08$~dex (depending on the particular OH line).

The OH lines investigated in this work are weak and situated on the linear part of the curve of growth. It makes the spectral lines under consideration insensitive to the choice of the microturbulence velocity value. Changing the microturbulence velocity by $1\, {\rm km\,s^{-1}}$ from the values listed in Table~\ref{table:obs-star} would change 3D--1D abundance correction and the average oxygen abundance by less than 0.01~dex.

\section{Discussion \label{sect:discussion}}

Abundance corrections provided in Tables~\ref{table:dabu-hd122563}--\ref{table:dabu-cs31082} clearly show that horizontal temperature fluctuations caused by convective motions in the stellar atmosphere have significantly larger impact on the strengths of IR~OH lines than the influence of the different temperature stratifications in the average $\xtmean{\mbox{3D}}$ and 1D model atmospheres. This is indicated by the different size of $\Delta_{\rm 3D-\langle3D\rangle}$ and $\Delta_{\rm \langle3D\rangle-1D}$ abundance corrections; the former is always larger than the latter (i.e. in absolute value). The physical reasons behind such behaviour are in fact rather obvious. As can be seen in Fig.~\ref{fig:temp-strat}, horizontal temperature fluctuations (as measured by $\Delta T_{\rm RMS} = \sqrt{\langle(T - T_0)^2\rangle_{x,y,t}}$, where $\langle . \rangle_{x,y,t}$ denotes temporal and horizontal averaging on surfaces of equal optical depth, while $T_0=\langle T \rangle_{x,y,t}$, is the depth-dependent average temperature) become increasingly larger in the outer atmosphere, reaching $\Delta T_{\rm RMS}\approx800$~K at $\log \tau_{\rm Ross}\approx-5$. Since IR~OH lines form rather high in the atmosphere, this leads to large non-linear horizontal variations of OH number densities and, thus, line opacities in the line forming regions. The result is that IR~OH lines become stronger in 3D, which in turn leads to negative (and relatively large) $\Delta_{\rm 3D-\langle3D\rangle}$ abundance corrections. On the other hand, the temperature difference between $\langle{\rm 3D}\rangle$ and 1D is small (and mostly positive) at all optical depths, which results in weaker lines in $\xtmean{\mbox{3D}}$ and, thus, small (and positive) $\Delta_{\rm \langle3D\rangle-1D}$ corrections.

In contrast, the influence of convection on the formation of the \ensuremath{\left[\ion{O}{i}\right]} oxygen line at 630~nm is very small. This is in line with our earlier findings \citep[see e.g.][]{DKS13} and is quite easily understood. Neutral oxygen is a majority species in the regions where the \ensuremath{\left[\ion{O}{i}\right]} 630~nm line forms, and because of its high ionization potential its abundance is insensitive to temperature fluctuations. At the same time, differences between the T profiles in the average $\xtmean{\mbox{3D}}$ and 1D models in the line formation regions are small (Fig.~\ref{fig:temp-strat}). This results in small $\Delta_{\rm 3D-\langle3D\rangle}$ and $\Delta_{\rm \langle3D\rangle-1D}$ abundance corrections, and thus, in small total correction, $\Delta_{\rm3D-1D}$. It is important to mention in this respect that the \ensuremath{\left[\ion{O}{i}\right]} line at 630~nm is equally insensitive to NLTE effects \citep[e.g.][and references therein]{A05}. One may therefore make a relatively safe assumption that 1D~LTE abundances obtained using the \ensuremath{\left[\ion{O}{i}\right]} 630~nm line are essentially unaffected by the 3D and NLTE effects, and may thus serve as an important reference point for validating oxygen abundances obtained from other lines. In this respect, we find it reassuring that in our EMP giants the 3D~LTE oxygen abundances obtained from IR~OH lines agree well with the abundances determined from the forbidden line (Table~\ref{table:o-abnd-tab}, Fig.~\ref{fig:ofe-evol}).

We also find that measuring the IR~OH lines in metal-poor giants is feasible at a resolution of 50\,000, provided a high S/N ratio of a few hundred can be obtained. The lines are sufficiently numerous so that a few of them are always likely to be free from contamination of telluric lines. In this respect the synthetic spectra of the terrestrial atmosphere computed by TAPAS \citep{BLF14} are precious for identifying the contaminated lines. In the present study we had enough clean lines and thus we felt it was not necessary to attempt to subtract the telluric components. Nevertheless, the fidelity of the TAPAS spectra is high enough so that subtraction of the telluric lines may be envisaged in cases in which a use of a larger number of lines is desirable.

Although our modelling of the formation of these lines is not entirely satisfactory, as can be deduced from the existence of slight trends of abundance with equivalent width and excitation potential shown in Figs.\,\ref{fig:o-abnd-1d} and \ref{fig:o-abnd-1d_exct}, respectively, we believe that the use of 3D model-atmospheres improves the modelling. We arrive at this conclusion by noting that on average 1D~LTE oxygen abundances from the IR OH lines are obviously too high with respect to what would be expected for the EMP stars in this \FeH\ range\footnote{The [O/Fe] abundance ratios from \citet[][]{CDS04} were obtained using solar oxygen abundance of $A({\rm O})=8.74$, whereas in our case we used $A({\rm O})=8.66$ from \citet[][]{AGS05}. To make the direct comparison possible, the [O/Fe] ratios from \citet[][]{CDS04} and the theoretical model of \citet[][]{FMC04} were shifted to the reference scale where the solar abundance is $A({\rm O})=8.66$.}. On the other hand, 3D~LTE abundances obtained from OH lines agree significantly better with the measurements obtained from \ensuremath{\left[\ion{O}{i}\right]} line than the 1D~LTE estimates do (Fig.~\ref{fig:ofe-evol}).

We note that similar trends of oxygen abundance with equivalent width were also noticed in the investigations of \citet{MB02} and \citet{BMS03} and they remain unexplained. We also investigated if the determined oxygen abundances depend on quantum number $J$, but could find no clear correlation. It is encouraging that, ignoring the trends, the line-to-line scatter of the abundances derived from the different lines is small enough so that the mean  can be considered a reasonable estimate of the oxygen abundance in a given star.

One can conclude that there are hints of a slight dependence of abundance on excitation potential in the case of BD-18:5550 (and perhaps also CS~31081-001; see Fig.~\ref{fig:o-abnd-1d_exct}). Nevertheless, in the case of BD-18:5550 the two lines with the highest excitation potential and the highest oxygen abundance also significantly deviate in the abundance--equivalent width plane (Fig.~\ref{fig:o-abnd-1d}). Why abundances obtained from these two lines differ is not clear; we note, however, that these lines are the weakest ones measured in the spectrum of this star.

We identify two ways to improve our modelling of these lines. The first is to increase the number of 3D model atmospheres used in the analysis. One clear limitation of the present investigation is the use of just one model atmosphere to describe the four programme stars, that, although they have been selected with very similar atmospheric parameters, span a range in surface gravities, effective temperatures, and metallicities. In particular, the sensitivity of the 3D corrections to the variations of atmospheric parameters needs to be investigated. Such an investigation is hampered by the long time needed to fully relax a 3D model atmosphere of a giant star.

The second way is the investigation of possible departures from the assumption of LTE, that has been made in the present investigation. Departures could occur both in the chemical equilibrium (formation and dissociation of OH molecules here are assumed to be an equilibrium process); and in the occupation number of different energy levels of the OH molecule (the Boltzmann law was assumed here).

In Fig.\,\ref{fig:ofe-evol} we show the [O/Fe] ratios in the four programme stars together with data taken from the literature based on the 630\,nm \ensuremath{\left[\ion{O}{i}\right]} line. The oxygen abundances are the ones we determined from the IR OH lines and the iron abundances have been taken from \citet{CDS04}. The points labelled 3D mean that only the oxygen abundance has been corrected for the 3D effect, but not the iron. In the case of iron, the 3D correction for such a model atmosphere should be around $-0.2$\,dex \citep{bonifacio09,DKS13}, although strongly dependent on excitation potential \citep{CAT07,collet09,DKS13}. On the other hand, we also know that \ion{Fe}{i} lines imply NLTE corrections of the same order of magnitude but in the opposite direction. In their comprehensive study of NLTE effects on iron in HD~122563 \citet{Mashonkina11} determined a NLTE correction of +0.16\,dex (for collisions with neutral hydrogen treated with the Drawin formalism and an efficiency S$_H=0.1$; neglecting the collisions raises the NLTE correction to +0.35). On the other hand, \citet{Mashonkina13} demonstrated that the 3D and NLTE corrections for iron cannot simply be added; this approach for HD~122563 leads to a discrepancy of $-0.75$\,dex between the abundance derived from the \ion{Fe}{i} and \ion{Fe}{ii} lines. The abundance of iron probably requires a full 3D-NLTE analysis, and we are currently unable to perform it. In this state of affairs we prefer to adopt the 1D~LTE result for iron.

With this choice, the 3D \OFe\ values obtained from OH lines appear to be in better agreement with the general trend defined by the abundances based on \ensuremath{\left[\ion{O}{i}\right]}. The average 3D~LTE oxygen abundance for the four stars obtained from OH lines is $\langle A({\rm O})\rangle = 6.61 \pm 0.21$, while that obtained from the 630~nm line is $\langle A({\rm O})\rangle = 6.52 \pm 0.33$. We note that in 1D~LTE we obtained $\langle A({\rm O})\rangle = 6.86 \pm 0.22$ from OH lines; errors in all cases here indicate the scatter due to star-to-star abundance variations.

The results shown in Fig.~\ref{fig:ofe-evol} are, by and large, consistent with those derived from the OH~UV lines by \citet{JGH10} in metal poor dwarfs, although these authors have adopted the strategy of adding 3D corrections and 1D~NLTE correction for \ion{Fe}{i}. This means that a direct comparison of the two results is not possible. Nevertheless, the results of \citet[][see their Fig.~14]{JGH10} show an average $\OFe = 0.6-0.8$, corrected for 3D and NLTE effects at $\FeH = -3.0$. In our case, we obtain an average $\OFe = 0.7$, corrected only for 3D effects, at $\FeH = -3.0$. On the other hand, the results of the two studies agree well with theoretical model of \citet{FMC04} which predicts $\OFe = 0.8$ for the same metallicity (in this comparison, [O/Fe] values from \citealt{JGH10} and \citealt{FMC04} were recomputed to the \citealt{AGS05} solar oxygen abundance scale).  In the future we plan to investigate the consistency of abundances obtained from IR~OH and UV lines in the same EMP stars.

\section{Conclusions}

The results obtained in our pilot project have demonstrated that infrared OH vibrational-rotational lines could be used successfully to determine oxygen abundance in EMP giants. This, however, requires high-resolution spectra obtained with $S/N$ ratios of at least a few hundred. Fortunately, IR~OH lines are significantly stronger than the \ensuremath{\left[\ion{O}{i}\right]} 630~nm line and thus can be more easily measured. In addition, quite a large number of IR~OH lines unaffected by telluric lines are typically available for the measurements, which makes the statistical significance of the determined oxygen abundance higher.

To determine oxygen abundances from IR~OH lines in EMP stars, however, the use of 3D hydrodynamical model atmospheres is imperative because these lines form in the outer atmosphere where horizontal temperature fluctuations (caused by convective motions) are large, which leads to large variations in line opacities. This results in significantly different line strengths predicted with the 3D hydrodynamical and 1D hydrostatic model atmospheres, with the resulting 3D--1D differences in oxygen abundances of $-0.2 \dots -0.3$~dex. When corrected for these effects with the aid of 3D hydrodynamical model atmospheres, oxygen abundances obtained from IR~OH lines generally agree well with those determined using the \ensuremath{\left[\ion{O}{i}\right]} 630~nm line, which is insensitive to either 3D or NLTE effects. Obviously, the role of NLTE effects in the IR~OH line formation still needs to be clarified, as well as the combined 3D NLTE effects for iron, something that is anticipated from the forthcoming studies.

\begin{acknowledgements}

We thank Jonas Klevas and Dainius Prakapavi\v{c}ius for their contribution during various stages of the paper preparation. This work was supported by grants from the Research Council of Lithuania (MIP-065/2013) and the bilateral French-Lithuanian programme ``Gilibert'' (TAP~LZ~06/2013, Research Council of Lithuania; 28471NE, Campus France). E.C. is grateful to the FONDATION MERAC for funding her fellowship. H.G.L. acknowledges financial support by the Sonderforschungsbereich SFB 881 ``The Milky Way System'' (subproject A4 and A5) of the German Research Foundation (DFG). We also thank the staff of ESO VLT for performing observations in service mode.

\end{acknowledgements}

\bibliographystyle{aa}

\begin{thebibliography}{}

\bibitem[{Akerman} {et al.}(2004)]{ACN04}
Akerman, C.~J., Carigi, L., Nissen, P.~E., Pettini, M., Asplund, M.
2004, \aap, 414, 931

\bibitem[Arnett(1996)]{Arnett} Arnett, D.\ 1996, Supernovae
and Nucleosynthesis by David Arnett.~Princeton University Press,
1996.~ISBN: 978-0-691-01147-9,

\bibitem[{Asplund} (2005)]{A05}
Asplund, M.
2005, \araa, 43, 481

\bibitem[{Asplund} {et al.}(2005)]{AGS05}
Asplund, M., Grevesse, N., \& Sauval, A. J.
2005, ASPC, 336, 25

\bibitem[{Barbuy} (1988)]{barbuy88}
Barbuy, B.
1988, \aap, 191, 121

\bibitem[{Barbuy \& Erdelyi-Mendes} (1989)]{BE89}
Barbuy, B., Erdelyi-Mendes, M.
1989, \aap, 214, 239

\bibitem[{Barbuy} {et al.}(2003)]{BMS03}
Barbuy, B., Mel\'{e}ndez, J., Spite, M., Spite, F., Depagne, E., et al.
2003, \apj, 588, 1072

\bibitem[{Bertaux} {et al.}(2014)]{BLF14}
Bertaux, J. L., Lallement, R., Ferron, S., Boonne, C., \& Bodichon, R.
2014, \aap, 564, 46

\bibitem[{Boesgaard} {et al.}(1999)]{BKD99}
Boesgaard, A.~M., King, J.~R., Deliyannis, C.~P., Vogt, S.~S.
1999, \aj, 117, 492

\bibitem[{Bonifacio} {et al.}(2009)]{bonifacio09}
Bonifacio, P., Spite, M., Cayrel, R., et al.
2009, \aap, 501, 519

\bibitem[{Caffau \& Ludwig} (2007)]{CL07}
Caffau, E., \& Ludwig, H.-G.
2007, \aap, 467, L11

\bibitem[{Caffau} {et al.}(2008)]{CLS08}
Caffau,~E., Ludwig,~H.-G., Steffen,~M., Ayres,~T.~R., Bonifacio,~P., Cayrel,~R., Freytag,~B., \& Plez,~B.
2008, \aap, 488, 1031

\bibitem[{Castelli} {et al.}(2003)]{castelli03}
Castelli, F. \&  Kurucz, R.~L.,
2003, Proceed. of IAU Symp. 210, Modeling of Stellar Atmospheres, eds. N. Piskunov et al., poster A20 on the enclosed CD-ROM

\bibitem[{Cayrel} {et al.}(2001)]{CAB01}
Cayrel, R., Andersen, J., Barbuy, B., Beers, T. C., Bonifacio, P., et al.
2001, NewAR, 45, 533

\bibitem[{Cayrel} {et al.}(2004)]{CDS04}
Cayrel, R., Depagne, E., Spite, M., Hill, V., Spite, F., et al.
2004, \aap, 416, 1117

\bibitem[{Collet} {et al.}(2007)]{CAT07}
Collet, R., Asplund, M., \& Trampedach, R.
2007, A\&A, 469, 687

\bibitem[Collet et al.(2009)]{collet09}
Collet, R., Nordlund, {\AA}., Asplund, M., Hayek, W., \& Trampedach, R.
2009, \memsai, 80, 719

\bibitem[Coxon
\& Foster(1982)]{coxonfoster} Coxon, J.~A., \& Foster, S.~C.\ 1982, Journal of Molecular Spectroscopy, 91, 243

\bibitem[{Creevey} {et al.}(2012)]{CTB12} 
Creevey, O.~L., Th\'{e}venin, F., Boyajian, T.~S., Kervella, P., Chiavassa, A., et al.
2012, \aap, 545, 17

\bibitem[{Cowan} {et al.}(2002)]{CSB02}
Cowan, J. J., Sneden, C., Burles, S., Ivans, I. I., Beers, T. C., et al.
2002, \apj, 572, 861

\bibitem[{Dobrovolskas} {et al.}(2013)]{DKS13}
Dobrovolskas,~V., Ku\v{c}inskas,~A., Steffen,~M., Ludwig,~H.-G., Prakapavi\v{c}ius,~D., et al.
2013, \aap, 559, A102

\bibitem[{Fran\c{c}ois} {et al.}(2004)]{FMC04}
Fran\c{c}ois, P., Matteucci, F., Cayrel, R., Spite, M., Spite, F., \& Chiappini, C.
2004, \aap, 421, 613

\bibitem[{Freytag} {et al.}(2012)]{FSL12}
Freytag, B., Steffen, M., Ludwig, H.-G., et al.
2012, J. Comp. Phys., 231, 919

\bibitem[{Fulbright \& Kraft}(1999)]{FK99}
Fulbright, J. P., \& Kraft, R. P.
1999, \aj, 118, 527

\bibitem[{Galazutdinov}(1992)]{galazutdinov}
Galazutdinov G. A.
1992, Special Astrophysical Observatory Preprint, No. 92, 2

\bibitem[Garc{\'{\i}}a P{\'e}rez et
al.(2006)]{ANA06} Garc{\'{\i}}a P{\'e}rez, A.~E., Asplund, M., Primas, F., Nissen, P.~E., \& Gustafsson, B.\ 2006, \aap, 451, 621

\bibitem[{Gonz{\'a}lez Hern{\'a}ndez} {et al.}(2010)]{JGH10}
Gonz{\'a}lez Hern{\'a}ndez, J.~I., Bonifacio, P., Ludwig, H.-G., et al.
2010, \aap, 519, A46

\bibitem[{Gratton \& Ortolani} (1986)]{GO86}
Gratton, R. G., \& Ortolani, S.
1986, \aap, 169, 201

\bibitem[{Grevesse \& Sauval} (1998)]{GS98}
Grevesse, N., \& Sauval, A. J.
1998, Space Sci. Rev., 85, 161

\bibitem[{Heiter \& Eriksson} (2006)]{HE06}
Heiter,~U., Eriksson,~K.
2006, \aap, 452, 1039

\bibitem[{Israelian} {et al.}(2001)]{IRG01}
Israelian, G., Rebolo,~R., Garc\'{i}a~L., Ram\'{o}n J., Bonifacio, P., et al.
2001, \apj, 551, 833

\bibitem[{K\"{a}ufl} {et al.}(2004)]{KBB04}
K\"{a}ufl, H.U., Ballester, P., Biereichel, P., Delabre, B., Donaldson, R. et al.
2004, SPIE, 5492, 1218

\bibitem[{Kraft} {et al.}(1992)]{KSL92}
Kraft, R. P., Sneden, C., Langer, G. E., \& Prosser, C. F.
1992, \aj, 104, 645

\bibitem[{Ku\v{c}inskas} {et al.}(2013)]{KSL13}
Ku\v{c}inskas,~A., Steffen,~M., Ludwig,~H.-G., Dobrovolskas,~V., Ivanauskas,~A. et al.
2013, \aap, 549, A14

\bibitem[{Kurucz}(1993)]{kurucz93}
Kurucz, R.
1993, ATLAS9 Stellar Atmosphere Programs and 2 km/s grid. Kurucz CD-ROM No. 13. Cambridge, Mass.: Smithsonian Astrophysical Observatory, 1993., 13

\bibitem[{Ludwig}(1992)]{ludwig92}
Ludwig, H.-G.
1992, Ph.D. Thesis, Univ. Kiel

\bibitem[Ludwig \& Ku\v{c}inskas(2012)]{LK12}
Ludwig,~H.-G., \& Ku\v{c}inskas,~A.
2012, \aap, 547, A118

\bibitem[{Ludwig} {et al.}(1994)]{LJS94}
Ludwig, H.-G., Jordan, S., \& Steffen, M.
1994, \aap, 284, 105

\bibitem[Matteucci(2012)]{Matteucci} Matteucci, F.\ 2012,
Chemical Evolution of Galaxies: , Astronomy and Astrophysics Library.~ISBN
978-3-642-22490-4.~Springer-Verlag Berlin Heidelberg, 2012,

\bibitem[{Mashonkina} {et al.}(2011)]{Mashonkina11}
Mashonkina, L., Gehren, T., Shi, J.-R., Korn, A.~J., \& Grupp, F.
2011, \aap, 528, A87

\bibitem[{Mashonkina} {et al.}(2013)]{Mashonkina13}
Mashonkina, L., Ludwig, H.-G., Korn, A., Sitnova, T., \& Caffau, E.
2013, \memsai\ Suppl., 24, 120

\bibitem[Mel{\'e}ndez et al.(2001)]{MBS01} Mel{\'e}ndez, J.,
Barbuy, B., \& Spite, F.\ 2001, \apj, 556, 858

\bibitem[Mel{\'e}ndez
\& Barbuy(2002)]{MB02} Mel{\'e}ndez, J., \& Barbuy, B.\ 2002, \apj, 575, 474

\bibitem[{Nordlund}(1982)]{nordlund82}
Nordlund, \AA.
1982, \aap, 107, 1

\bibitem[{Sbordone} {et al.}(2004)]{sbordone04}
Sbordone, L., Bonifacio, P., Castelli, F., \& Kurucz, R. L.
2004, \memsai, 5, 93

\bibitem[{Sbordone}(2005)]{sbordone05}
Sbordone, L.
2005, \memsai, 8, 61

\bibitem[{Sneden} {et al.}(1991)]{SKP91}
Sneden, C., Kraft, R. P., Prosser, C. F., \& Langer, G. E.
1991, \aj, 102, 2001

\bibitem[{Spite} {et al.}(2005)]{SCP05}
Spite, M., Cayrel, R., Plez, B., Hill, V., Spite, F., et al.
2005, \aap, 430, 655

\bibitem[{Takeda \& Takada-Hidai} (2013)]{takeda13}
Takeda, Y. \& Takada-Hidai, M.
2013, PASJ, 65, 65

\bibitem[{V\"{o}gler} {et al.}(2004)]{VBS04}
V\"{o}gler, A., Bruls, J. H. M. J., \& Sch\"{u}ssler, M.
2004, \aap, 421, 741

\end{thebibliography}

\Online

\begin{appendix}

\section{List of infrared OH lines for oxygen abundance determination}

In this work we used a number of vibrational-rotational lines ($X^{-2}\Pi$) from the first-overtone sequence ($\Delta v = 2$) located in the spectral range between $1514-1548$ and $1595-1632$~nm. The atomic parameters of the IR~OH lines and their equivalent widths measured in the target EMP giants and also the derived oxygen abundances are provided in Table~\ref{table:line-list}.

\begin{table*}[tbh]
\caption{The list of the OH spectral lines used in the abundance analysis.}
\label{table:line-list}
\centering
\begin{tabular}{cccc c@{}c@{}c@{}c@{}c@{}c@{}c@{}c@{}}
\hline\hline
$\lambda$,   & $\chi$, & $\log gf$ &$v'-v''$ & \multicolumn{2}{c}{HD 122563}  & \multicolumn{2}{c}{HD 186478}   &  \multicolumn{2}{c}{BD-18:5550}  & \multicolumn{2}{c}{CS 31082-001}   \\
nm           & eV      &           &      & $W$,    & $A$(O) & $W$,      & $A$(O) &  $W$,     & $A$(O) & $W$,  & $A$(O) \\
             &         &           &      & (pm)    &      &(pm)       &      &  (pm)       &      & (pm)      &      \\
\hline
1514.5770    & 0.164   & $-5.447$  & 0-2 &   \ldots  & \ldots &  1.52    & 7.15   & 0.52      & 6.72   & \ldots  & \ldots \\
1514.7939    & 0.164   & $-5.447$  & 0-2 &   1.20    & 6.61   &  1.71    & 7.20   & 0.56      & 6.76   & \ldots  & \ldots \\
1526.4604    & 0.210   & $-5.428$  & 0-2 &   \ldots  & \ldots &  1.13    & 7.04   & \ldots     & \ldots & \ldots  & \ldots \\
1526.6167    & 0.210   & $-5.428$  & 0-2 &   \ldots  & \ldots &  1.46    & 7.16   & \ldots     & \ldots & 0.82    & 7.09   \\
1527.8524    & 0.205   & $-5.382$  & 0-2 &   1.30    & 6.62   &  1.60    & 7.15   & 0.56      & 6.72   & 0.89    & 7.07   \\
1528.1052    & 0.205   & $-5.382$  & 0-2 &   1.38    & 6.64   &  1.53    & 7.12   & 0.64      & 6.78   & \ldots  & \ldots \\
1541.9460    & 0.250   & $-5.323$  & 0-2 &   \ldots  & \ldots &  1.69    & 7.15   & \ldots     & \ldots & \ldots  & \ldots \\
1542.2366    & 0.250   & $-5.323$  & 0-2 &   \ldots  & \ldots &  1.87    & 7.06   & \ldots     & \ldots & \ldots  & \ldots \\
1605.2765    & 0.639   & $-4.910$  & 1-3 &   \ldots  & \ldots &  1.44    & 7.09   & \ldots     & \ldots & \ldots  & \ldots \\
1605.5464    & 0.640   & $-4.910$  & 1-3 &   1.09    & 6.54   &  \ldots  & \ldots & \ldots     & \ldots & \ldots  & \ldots \\
1606.1700    & 0.476   & $-5.159$  & 0-2 &   \ldots  & \ldots &  1.39    & 7.11   & 0.39      & 6.60   & \ldots  & \ldots \\
1606.9524    & 0.472   & $-5.128$  & 0-2 &   \ldots  & \ldots &  \ldots  & \ldots & 0.36      & 6.53   & \ldots  & \ldots \\
1619.0132    & 0.688   & $-4.893$  & 1-3 &   1.71    & 6.57   &  \ldots  & \ldots & \ldots     & \ldots & 0.97    & 6.91   \\
1619.2130    & 0.688   & $-4.893$  & 1-3 &   0.97    & 6.52   &  \ldots  & \ldots & \ldots     & \ldots & \ldots  & \ldots \\
1620.4076    & 0.683   & $-4.851$  & 1-3 &   1.84    & 6.69   &  \ldots  & \ldots & \ldots     & \ldots & 0.77    & 6.84   \\
1623.0465    & 0.905   & $-5.059$  & 2-4 &   \ldots  & \ldots &  \ldots  & \ldots & 0.35      & 6.81   & \ldots  & \ldots \\
1623.1419    & 0.905   & $-5.059$  & 2-4 &   \ldots  & \ldots &  \ldots  & \ldots & 0.35      & 6.90   & \ldots  & \ldots \\
1625.1660    & 0.543   & $-5.115$  & 0-2 &   \ldots  & \ldots &  0.93    & 6.96   &  \ldots    & \ldots & \ldots  & \ldots \\
1625.5019    & 0.538   & $-5.087$  & 0-2 &   1.61    & 6.75   &  1.61    & 7.17   & \ldots     & \ldots & \ldots  & \ldots \\
1626.0155    & 0.539   & $-5.087$  & 0-2 &   1.63    & 6.75   &  1.22    & 7.04   & \ldots     & \ldots & \ldots  & \ldots \\
\hline
\end{tabular}
\end{table*}

\section{3D--1D abundance corrections for IR~OH lines and the forbidden oxygen line}

In Tables \ref{table:dabu-hd122563}-\ref{table:dabu-cs31082} we provide 3D--1D abundance corrections that were computed for all OH lines and forbidden oxygen line, in each EMP giant studied here.

\begin{table}[tb]
\caption{Abundance corrections for HD~122563.}
\label{table:dabu-hd122563}
\centering
\begin{tabular}{ccccc}
\hline\hline
$\lambda$, &   $EW$, &   ${\rm 3D}-\xtmean{\rm 3D}$  &  $\xtmean{\rm 3D}-{\rm 1D}$   &  3D--1D    \\
nm         &  (pm)   &  (dex)     & (dex)      & (dex)     \\
\hline
\multicolumn{5}{l}{\ensuremath{\left[\ion{O}{i}\right]} line}                                   \\
630.0.304  &  $0.64^{a}$ & $-0.044$ & 0.039  &  $-0.005$ \\
\multicolumn{5}{l}{OH lines}                                     \\
1514.7939  &  1.20   &  $-0.41$  &   0.13    &  $-0.29$ \\
1527.8524  &  1.30   &  $-0.41$  &   0.12    &  $-0.28$ \\
1528.1052  &  1.38   &  $-0.41$  &   0.12    &  $-0.28$ \\
1605.5464  &  1.09   &  $-0.34$  &   0.12    &  $-0.21$ \\
1619.0132  &  1.71   &  $-0.33$  &   0.12    &  $-0.21$ \\
1619.2130  &  0.97   &  $-0.33$  &   0.12    &  $-0.21$ \\
1620.4076  &  1.84   &  $-0.33$  &   0.12    &  $-0.21$ \\
1625.5019  &  1.61   &  $-0.35$  &   0.12    &  $-0.23$ \\
1626.0155  &  1.63   &  $-0.35$  &   0.12    &  $-0.23$ \\
\hline
\end{tabular}
\begin{list}{}{}
\item[$^{\mathrm{a}}$] adopted from \citet{CDS04}
\end{list}
\end{table}

\begin{table}[tb]
\caption{Same as in Table~\ref{table:dabu-hd122563} but for HD~186478.}
\label{table:dabu-hd186478}
\centering
\begin{tabular}{ccccc}
\hline\hline
$\lambda$, &   $EW$, &   ${\rm 3D}-\xtmean{\rm 3D}$  &  $\xtmean{\rm 3D}-{\rm 1D}$   &  3D--1D    \\
nm         &  (pm)   &  (dex)     & (dex)      & (dex)     \\
\hline
\multicolumn{5}{l}{\ensuremath{\left[\ion{O}{i}\right]} line}                                   \\
630.0.304  &  $0.93^{a}$ & $-0.045$ &  0.040 &  $-0.005$ \\
\multicolumn{5}{l}{OH lines}                                     \\
1514.5770  &  1.52   &  $-0.42$  &   0.13    &  $-0.29$ \\
1514.7939  &  1.71   &  $-0.42$  &   0.12    &  $-0.29$ \\
1526.4604  &  1.13   &  $-0.41$  &   0.12    &  $-0.28$ \\
1526.6167  &  1.46   &  $-0.40$  &   0.12    &  $-0.28$ \\
1527.8524  &  1.60   &  $-0.41$  &   0.12    &  $-0.28$ \\
1528.1052  &  1.53   &  $-0.41$  &   0.12    &  $-0.28$ \\
1541.9460  &  1.69   &  $-0.40$  &   0.12    &  $-0.27$ \\
1542.2366  &  1.87   &  $-0.39$  &   0.12    &  $-0.27$ \\
1605.2765  &  1.44   &  $-0.34$  &   0.12    &  $-0.22$ \\
1606.1700  &  1.39   &  $-0.36$  &   0.12    &  $-0.24$ \\
1625.1660  &  0.93   &  $-0.35$  &   0.12    &  $-0.23$ \\
1625.5019  &  1.61   &  $-0.35$  &   0.12    &  $-0.23$ \\
1626.0155  &  1.22   &  $-0.35$  &   0.12    &  $-0.23$ \\
\hline
\end{tabular}
\begin{list}{}{}
\item[$^{\mathrm{a}}$] adopted from \citet{CDS04}
\end{list}
\end{table}

\begin{table}[tb]
\caption{Same as in Table~\ref{table:dabu-hd122563} but for BD--18:5550.}
\label{table:dabu-bd185550}
\centering
\begin{tabular}{ccccc}
\hline\hline
$\lambda$, & $EW$,  &   ${\rm 3D}-\xtmean{\rm 3D}$  &  $\xtmean{\rm 3D}-{\rm 1D}$   &  3D--1D    \\
nm         &(pm)    &  (dex)     & (dex)      & (dex)     \\
\hline
\multicolumn{5}{l}{\ensuremath{\left[\ion{O}{i}\right]} line}                                   \\
630.0.304  &  $0.15^{a}$ & $-0.043$ & 0.038   &  $-0.006$ \\
\multicolumn{5}{l}{OH lines}                                     \\
1514.5770  &  0.515   &  $-0.41$  &   0.12    &  $-0.29$ \\
1514.7939  &  0.562   &  $-0.41$  &   0.12    &  $-0.29$ \\
1527.8524  &  0.560   &  $-0.41$  &   0.12    &  $-0.28$ \\
1528.1052  &  0.636   &  $-0.41$  &   0.12    &  $-0.28$ \\
1606.1700  &  0.390   &  $-0.36$  &   0.12    &  $-0.24$ \\
1606.9524  &  0.361   &  $-0.36$  &   0.12    &  $-0.24$ \\
1623.0465  &  0.352   &  $-0.30$  &   0.12    &  $-0.18$ \\
1623.1419  &  0.349   &  $-0.30$  &   0.12    &  $-0.18$ \\
\hline
\end{tabular}
\begin{list}{}{}
\item[$^{\mathrm{a}}$] adopted from \citet{CDS04}
\end{list}
\end{table}

\begin{table}[tb]
\caption{Same as in Table~\ref{table:dabu-hd122563} but for CS~31082-001.}
\label{table:dabu-cs31082}
\centering
\begin{tabular}{ccccc}
\hline\hline
$\lambda$, & $EW$,  &   ${\rm 3D}-\xtmean{\rm 3D}$  &  $\xtmean{\rm 3D}-{\rm 1D}$   &  3D--1D    \\
nm         &(pm)    &  (dex)     & (dex)      & (dex)     \\
\hline
\multicolumn{5}{l}{\ensuremath{\left[\ion{O}{i}\right]} line}                                   \\
630.0.304  &  $0.27^{a}$ & $-0.044$ & 0.038   &  $-0.006$ \\
\multicolumn{5}{l}{OH lines}                                     \\
1526.6167  &  0.82    &  $-0.41$  &   0.12    &  $-0.28$ \\
1527.8524  &  0.89    &  $-0.41$  &   0.12    &  $-0.28$ \\
1619.0132  &  0.97    &  $-0.33$  &   0.12    &  $-0.21$ \\
1620.4076  &  0.77    &  $-0.33$  &   0.12    &  $-0.21$ \\
\hline
\end{tabular}
\begin{list}{}{}
\item[$^{\mathrm{a}}$] adopted from \citet{CDS04}
\end{list}
\end{table}

\begin{figure*}[tb]
\centering
\includegraphics[width=\textwidth]{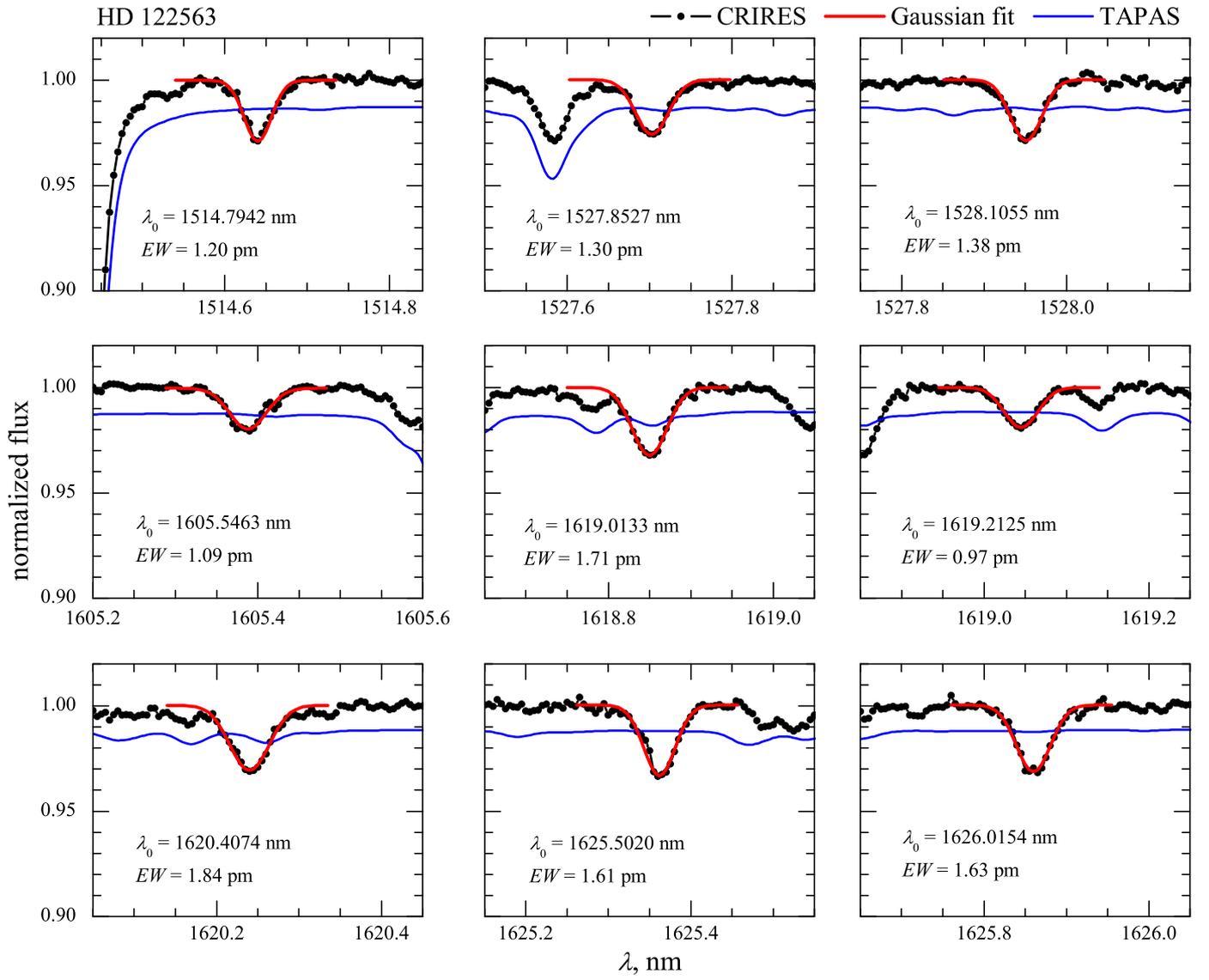}
\caption{Excerpts from CRIRES spectra centred on IR~OH lines in HD~122563 (black dots). Gaussian best-fits to the observed IR~OH line profiles are shown as red lines. Blue solid lines show the synthetic telluric spectrum computed with the TAPAS online service \citep{BLF14} which is shifted downwards by 0.01 for better readability. Rest frame wavelengths ($\lambda_0$) and measured equivalent width values ($EW$) of the IR~OH lines are provided at the bottom of each panel.}
\label{fig:hd122563-spect}
\end{figure*}

\begin{figure*}[tb]
\centering
\includegraphics[width=\textwidth]{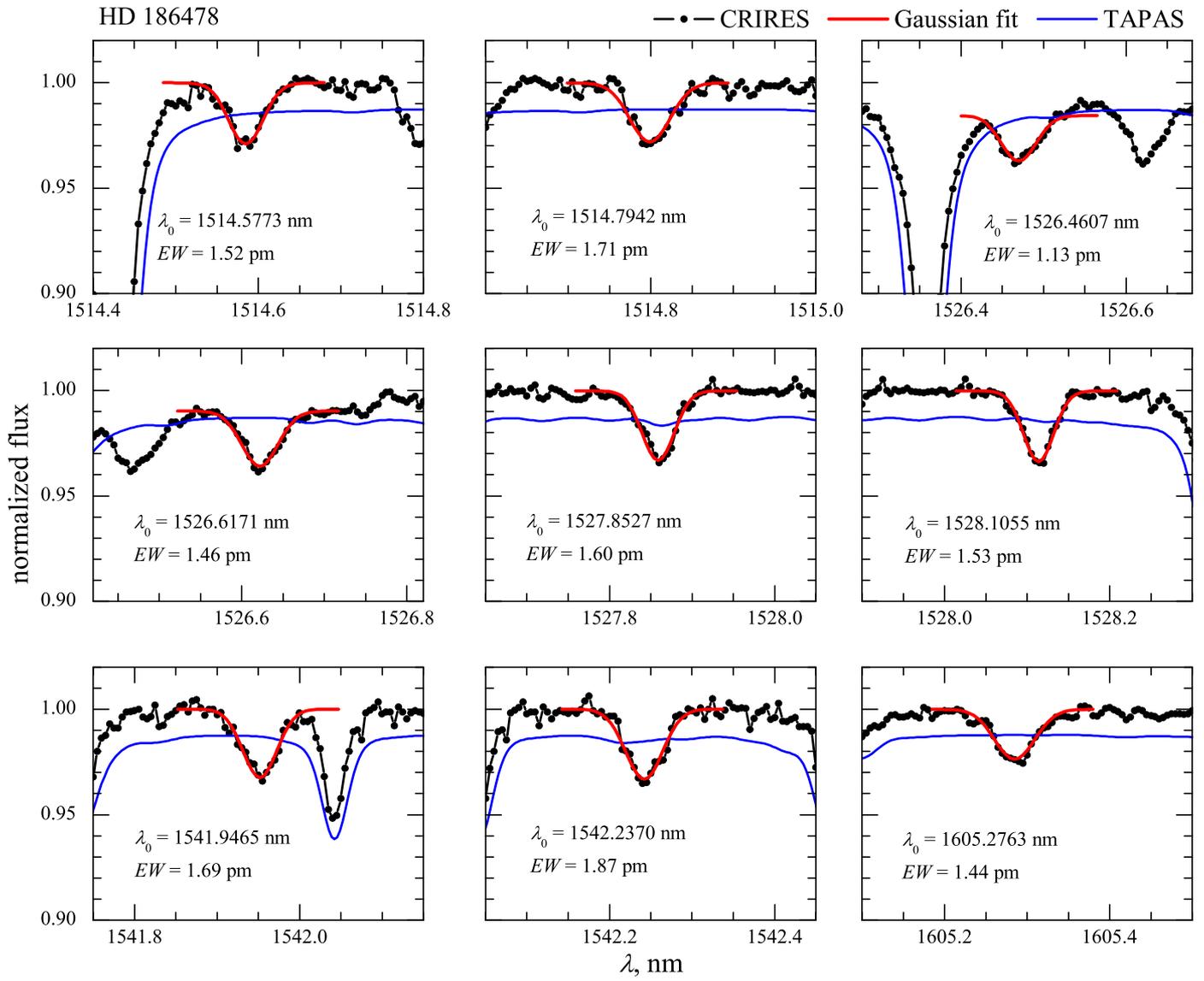}
\caption{The same as in Fig.~\ref{fig:hd122563-spect} but for HD~186478.}
\label{fig:hd186478-spect-a}
\end{figure*}

\begin{figure*}[tb]
\centering
\includegraphics[width=\textwidth]{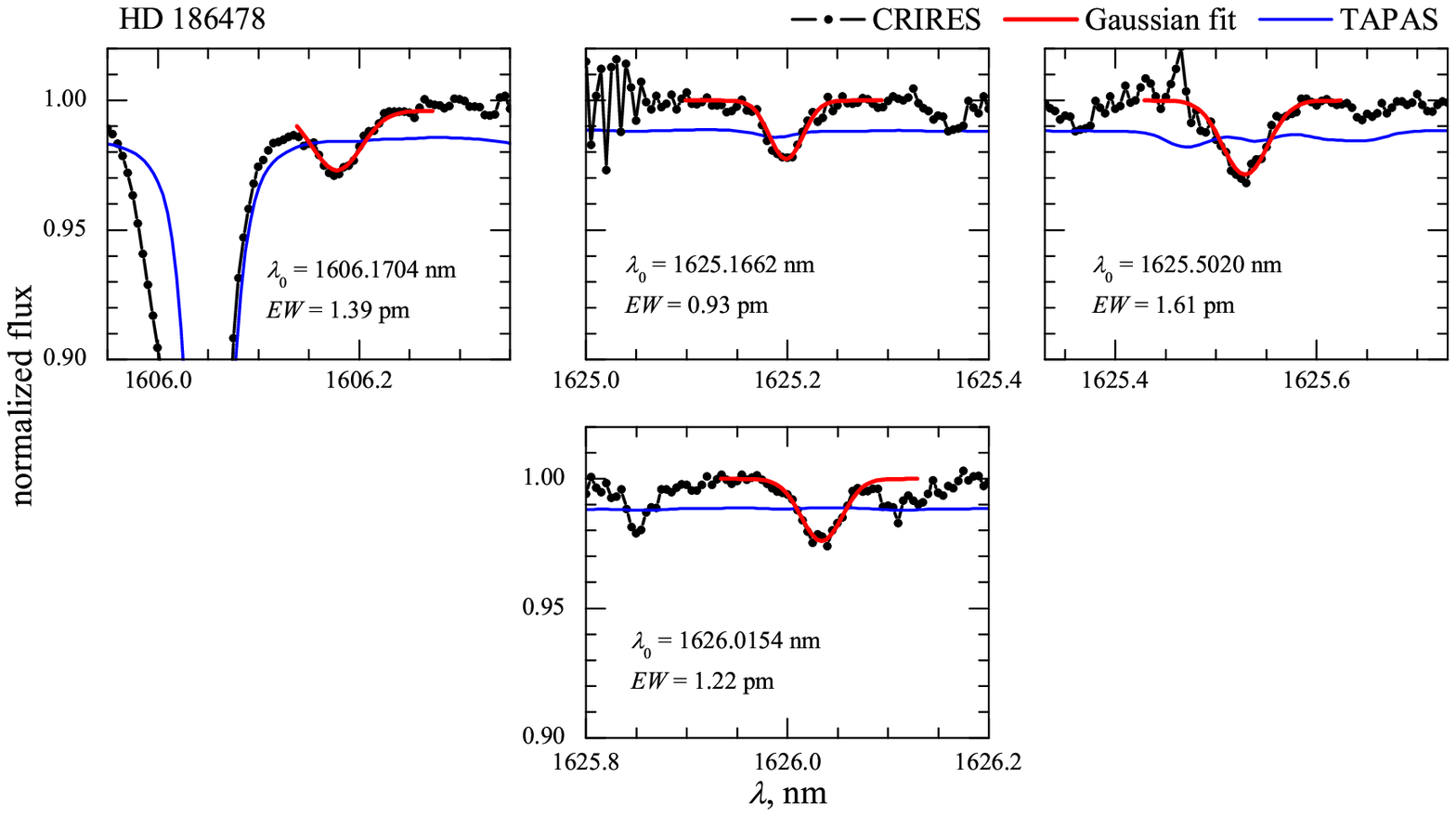}
\caption{The same as in Fig.~\ref{fig:hd186478-spect-a}. }
\label{fig:hd186478-spect-b}
\end{figure*}

\begin{figure*}[tb]
\centering
\includegraphics[width=\textwidth]{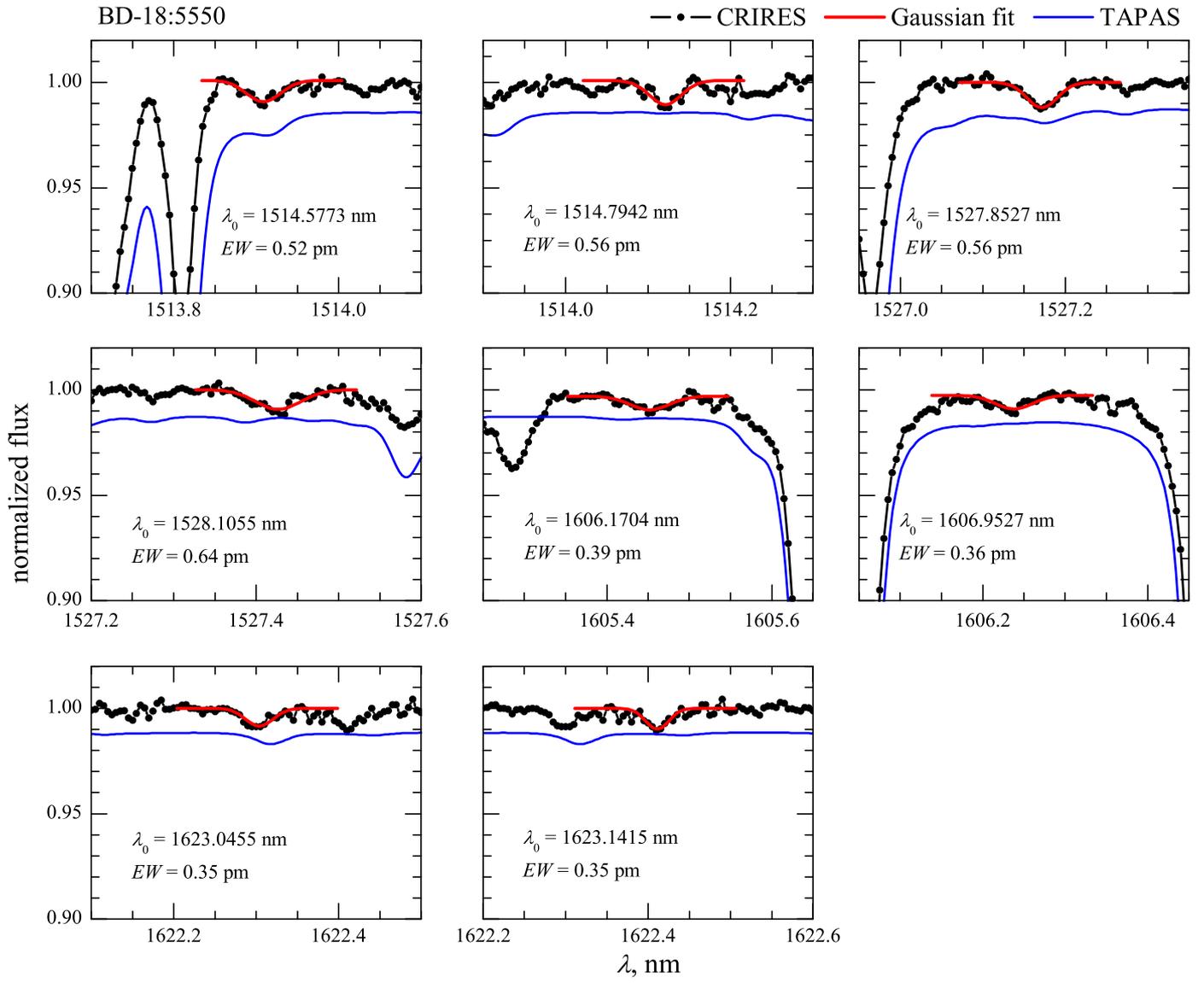}
\caption{The same as in Fig.~\ref{fig:hd122563-spect} but for BD~--18:5550.}
\label{fig:bd185550-spect}
\end{figure*}

\begin{figure*}[tb]
\centering
\includegraphics[width=\textwidth]{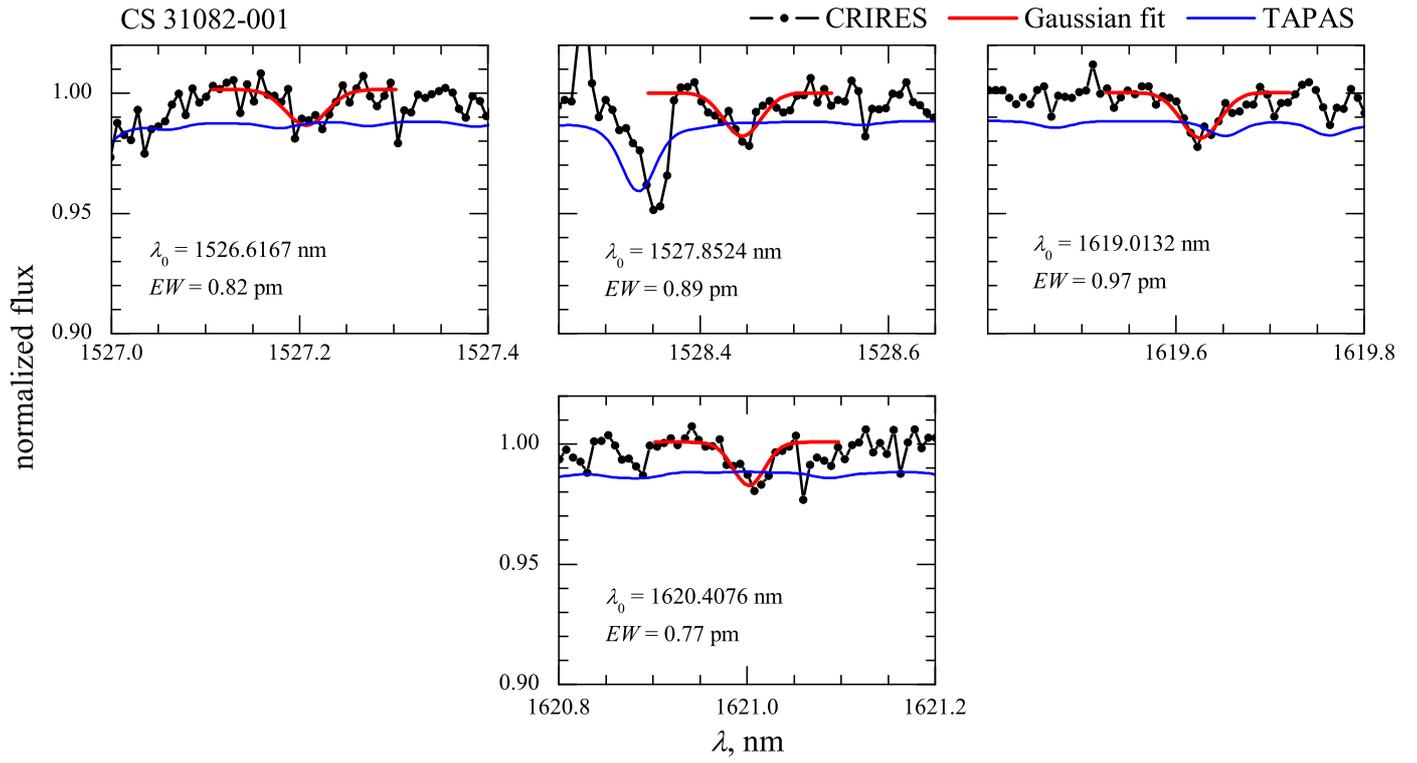}
\caption{The same as in Fig.~\ref{fig:hd122563-spect} but for CS~31082-001.}
\label{fig:cs31082-spect}
\end{figure*}

\end{appendix}

\end{document}